\documentclass[onecolumn, 12pt, journal, draftclsnofoot]{IEEEtran} 

\usepackage{mathptmx}      % use Times fonts if available on your TeX system
\usepackage{latexsym}
\usepackage{graphicx}
\usepackage[english]{babel}

\usepackage{amsmath,amssymb,mathrsfs,amsfonts,dsfont}
\usepackage{algorithmic}
\usepackage{balance}
\usepackage{cite}
\usepackage{fixmath}
\usepackage{multirow}
\usepackage{graphics}
\usepackage{color}
\usepackage{textcomp}
\usepackage{mathtools}
\usepackage{latexsym}
\usepackage{graphicx}
\usepackage{caption}
\usepackage[english]{babel}
\usepackage{amsmath,amssymb,mathrsfs,amsfonts,dsfont}
\usepackage{algorithmic}
\usepackage{hyperref}
\usepackage{cleveref}
\usepackage{fixmath}
\usepackage{multirow}
\usepackage{graphics}
\usepackage{color}
\usepackage{textcomp}
\usepackage{mathtools}

\newcommand{\site}{\boldsymbol{s}}
\newcommand\myeq{\stackrel{\mathclap{\normalfont\mbox{(a)}}}{=}}

\begin{document}

	\title{Dynamic-TDD Interference Tractability Approaches and Performance Analysis in Macro-Cell and Small-Cell Deployments}
	
	\author{Jalal Rachad, 
		Ridha Nasri and
		Laurent Decreusefond
		
		\thanks{Jalal Rachad and Ridha Nasri are with Orange Labs, 40-48 Avenue de la R\'epublique, 92320 Ch\^atillon, France. Email: jalal.rachad@ieee.org, ridha.nasri@orange.com}% <-this % stops a space
		\thanks{Jalal Rachad and Laurent Decreusefond are with LTCI Telecom Paris-Tech, Institut Polytechnique de Paris, Paris, France. Email: jalal.rachad@telecom-paristech.org, laurent.decreusefond@mines-telecom.fr}
		%\thanks{This work has been accepted in part in IEEE VTC 2018, Porto, Portugal.}% <-this % stops a space
	}

	%\markboth{Journal of \LaTeX\ Class Files,~Vol.~14, No.~8, August~2018}%

	\maketitle
	
	\begin{abstract}
	Meeting the continued growth in data traffic volume, Dynamic Time Division Duplex (D-TDD) has been introduced as a solution to deal with the uplink (UL) and downlink (DL) traffic asymmetry, mainly observed for dense heterogeneous network deployments, since it is based on instantaneous traffic estimation and provide more flexibility in resource assignment. However, the use of this feature requires new interference mitigation schemes capable to handle two additional types of interference between cells in opposite transmission direction: DL to UL and UL to DL interference. The aim of this work is to provide a complete analytical approach to model inter-cell interference in macro-cell and dense small-cell networks. We derive the explicit expressions of Interference to Signal Ratio (ISR) at each position of the network, in both DL and UL, to quantify the impact of each type of interference on the system performance. Also, we provide the explicit expressions of the coverage probability as  functions of different system parameters by covering different scenarios. Finally, through system level simulations, we analyze the feasibility of D-TDD implementation in both deployments and we compare its performance to the static-TDD (S-TDD) configuration.
	\end{abstract}
	

	\IEEEpeerreviewmaketitle

	\section{Introduction}
	
	\subsection{Context of the study}
	
	\IEEEPARstart{A}{s} the number of mobile users and the volume of data traffic are expected to continue increasing in the upcoming years, future mobile cellular networks need to support this proliferation through upgrading their features and key technologies. The upcoming fifth generation (5G) of cellular network radio interface, known as New Radio (NR), is being designed by considering flexibility in features definition in order to satisfy diverse use cases with different users' requirements. D-TDD is expected to be one of the major keys of 5G NR. It has been proposed in order to deal with traffic asymmetry since it enables the dynamic adjustment of UL and DL resource assignment  according to the instantaneous traffic variations. However, D-TDD system is severely limited by a strong mutual interference between the UL and DL transmissions because those two directions share the same frequency band. Hence, two types of interference appear: DL to UL (impact of DL other cell interference on UL signal received by the studied cell) and UL to DL (impact of UL mobile users transmission, located in other cells, on DL signal received by a mobile user located in the studied cell). Those additional interference, mainly DL to UL, are usually more difficult to deal with because of the LOS (Line Of Sight) presence between highly elevated base stations (BSs) transmitting with high power level and also because the mobiles can move around randomly. Thus, this duplexing mode can be more convenient with heterogeneous networks (HetNets) as small-cells are considered well isolated from each others and also from the macro-cell layers.\\
	
	Furthermore, in order to mitigate interference in D-TDD system, 3GPP (3rd Generation Partnership Project) standard advices new approach for enhanced Interference Mitigation and Traffic Adaptation (eIMTA) in dynamic environment \cite{astely2013lte}. Cell clustering scheme is an efficient technique that can be used to deal with D-TDD interference. Cells that suffer from high DL to UL interference level between each others can be gathered in the same cluster and use the same UL-DL configuration. In the same time, transmission directions in different clusters can be dynamically adapted. This technique can be very efficient for HetNet when it is combined with enhanced Inter Cell Interference Coordination (eICIC), introduced in 3GPP Release 10,  or Further eICIC (FeICIC), introduced in 3GPP Release 11. eICIC and FeICIC are based on time domain partitioning: macro-cell BSs reduce their transmitted power level during some sub-frames called Almost Blank Sub-frames (ABS) so that small-cells can adjust UL-DL portions dynamically during those sub-frames according to the traffic variations.
	
	\subsection{Related works}
	
D-TDD has been widely investigated in the available scientific literature; see for instance \cite{jeong2002cochannel,holma2000interference,ji2013dynamic,pirinen2015challenges,nasreddine2016interference,zhong2015dynamic,elbamby2014dynamic,khoryaev2012performance,rachad2018interference,lukowa2019centralized,lukowa2018performance}. The first study dates back to 2002 with the work in \cite{jeong2002cochannel} where performance of a D-TDD fixed cellular network in UL transmission were investigated. Authors in \cite{jeong2002cochannel} proposed a time slot assignment method to improve the UL outage performance. In our recent work \cite{rachad2018interference}, we have proposed an analytical tractability approach to model interference generated by D-TDD in a macro-cell deployment. We have shown that D-TDD is only used in favor of DL transmission cycle. However, during the UL transmission cycle, DL to UL interference may cause a substantial performance degradation. To reduce the impact of DL transmission of other cells on the UL received signal, we have proposed a cell clustering scheme that somehow improves D-TDD system UL performance in a dense small-cells' network. Performance of D-TDD system were also investigated in \cite{yu2015dynamic} for a particular small-cells' architecture, known as phantom cells, in UL and DL transmission directions. For the analytical approach, authors of \cite{yu2015dynamic} used tools from stochastic geometry to model phantom cells and user locations in order to derive $SINR$ distributions in DL and UL. Also, an inter-cell interference coordination scheme has been proposed. Similarly in \cite{ding2018performance}, D-TDD has been analyzed considering a dense small-cell network. In \cite{sun2015d2d}, a two-tier Device to Device enhanced HetNet operating with D-TDD has been studied. Authors have proposed an analytical framework to evaluate the coverage probability and network throughput using stochastic geometry. Likewise in \cite{kulkarni2017performance}, authors have provided a comparison between static and dynamic TDD in millimeter wave (mm-wave) cellular network, in terms of $SINR$ distributions and mean rates, considering synchronized and unsynchronized access-backhaul. Additionally, in order to make D-TDD feasible, some interference mitigation techniques have been proposed in literature, such as cell clustering; see for instance \cite{gao2015performance,lin2015dynamic} and \cite{rachad2018interference}. It was discussed in \cite{lin2015dynamic} a soft reconfiguration method based on cell clustering so as to allow cells in the same cluster to change dynamically the UL/DL configuration but inter-cluster interference still exists. Also, several works have discussed radio resource management and optimization approaches to deal with cross slot interference generated by D-TDD \cite{lukowa2019centralized}.

On the other hand, inter-cell interference are the major issue that obstructs the achievement of high performance in terms of data rate and spectral efficiency, especially in dense cellular networks deployments. Interference tractability is mainly related to the network geometry modeling. It has been proposed in literature several approaches to tackle this problem in both UL and DL transmission directions. The most frequently used approaches adopt random models, in which BSs are distributed in the plane according to a Poisson Point Process (PPP) \cite{andrews2011tractable,novlan2013analytical,dhillon2012modeling}, and the regular hexagonal model \cite{nasri2016Analytical}. Each model has its own advantages and disadvantages. Fro and engineering point of view, it is always desirable to evaluate inter-cell interference received at each location taken by a mobile user, which cannot be accomplished by using random models where only the interference distribution is determined. However, as one of the major ways leading 5G networks implementation is network densification and small-cell deployments, it becomes knotty to model the topology with deterministic models. Therefore, stochastic geometry appears to be efficient to determine the relevant metrics required to analyze radio performance in terms of the probabilistic parameters of HetNets.

	\subsection{Contribution}
	
	The main contribution of this paper is to provide a complete analytical framework for interference tractability in macro-cell deployment and dense small-cell network. We model macro-cells by using a regular hexagonal network with infinite number of sites. We treat, in particular, the explicit evaluation of $ISR$ at each position taken by a mobile user in the network, in terms of convergent series, by covering the four types of interference generated in D-TDD based network. This metric is very useful for link budget tools in which the expression of the average perceived interference is required in each position. To model small-cells, we adopt the widely used spatial PPP and we show how to exploit the mathematical framework based on stochastic geometry and satisfy in the same time D-TDD assumptions. Additionally,  we derive the explicit expressions of the coverage probability (SINR distribution) for a typical cell in DL and UL for both macro-cell and small-cell networks. This metric is related to throughput distribution and it is useful for cell throughput dimensioning. Finally, we analyze, through system level simulations, performance  of D-TDD based network and its comparison with Static-TDD (S-TDD) considering different system parameters. 
	
	\subsection{Paper organization} 
	
	The rest of the paper is organized as follows: In Section 2, we describe the D-TDD model, the macro-cells' network,  the small-cells' deployment and the propagation model. In Section 3, we provide the analytical analysis regarding the explicit derivation of $ISR$ and the coverage probability in macro-cells' deployment. The theoretical analysis of small-cells' network performance is given in Section 4. It is important to note that Section 3 and 4 are independent. Simulation results are provided in Section 5. Section 6 concludes the paper.
	
	\section{System models and notations}
	
	\subsection{Dynamic TDD model}
	
	To model the D-TDD system, we assume that all cells initially operate synchronously in DL or UL. This setup can be considered as a baseline scenario characterizing performance of existing synchronous TDD systems i.e., S-TDD. After a period of time, it is assumed that all cells select randomly UL or DL frame portions based on traffic conditions. Four types of interference henceforth appear depending on the transmission direction: \textit{i}) when the serving cell transmits to a given mobile location, DL and UL interference effect on DL useful transmission appears; \textit{ii}) when the serving cell receives signals from mobiles,  UL and DL interference impact on UL transmission rises. It is considered hereafter that the scheduler does not allocate the same spectral resources to different mobile users in one cell at the same time (e.g., TD-LTE scheduling). So, intra-cell interference is not considered. Therefore in a given cell,  we consider that during a sub-frame of interest (i.e., when D-TDD is activated), there is one active transmission whether on DL or UL with full-buffer traffic model. To characterize the transmission directions of cells, denoted by $\site$, we consider two Bernoulli Random variables (RVs) $\chi_d(\site)$ and $\chi_u(\site)$ such that $\mathbb{P}(\chi_d(\site)=1)=\alpha_d$ and $\mathbb{P}(\chi_u(\site)=1)=\alpha_u$. $\chi_d(\site)$ refers to the DL transmission cycle of a cell $\site$ and $\chi_u(\site)$ refers to its UL transmission cycle during a D-TDD sub-frame. It is important to mention that a cell $\site$ cannot operate in DL and UL during the same TTI. Hence, to avoid this case, we add the following condition $\chi_u(\site)=1-\chi_d(\site)$. This means that $\alpha_{d}=1-\alpha_{u}$.

\subsection{Network models}

Interference in cellular networks  are the  major issue that obstructs the achievement of high performance in terms of data rate and spectral efficiency. Telecommunication actors continuously attempt to minimize it during all the phases of a technology conception, since it is related to network performance. In radio engineering, interference margin, known also as noise rise, is used to perform link budget tools. However, this notion does not describe the real perceived interference and does not take into account the geometry of the studied area that impacts in one way or another performance. Thus, the analytical tractability of interference is of prime importance. Having a tractable mathematical model can always give better results and avoid recourse to extensive simulations. \\

In effect, interference is related to the network geometry and the spatial distribution of users. Most considered models that can be found in literature are the deterministic models such as the  regular hexagonal network  and random models based on spatial point processes. Hexagonal network is the basic model for network design in radio engineering. It is considered effective for network having fixed cell radius such as macro-cell deployments. Nevertheless, this model is not useful to describe heterogeneous networks topology. Small-cells usually occupy unplanned random positions which makes stochastic point processes  practical to model their random distribution in dense urban environments. Homogeneous PPP (HPPP) is a very popular model in cellular networks in which BSs and mobile users spatial distribution are modeled according to independent PPPs. However, despite the popularity of HPPP and its tractability, this model can not fit with the geometry of real cellular networks because of the repulsive behavior of transmitting nodes. Also, with this model, one cannot evaluate interference at each arbitrary user location and only its distribution that can be determined \cite{nasri2016Analytical} and \cite{nasri2015tractable}. In the remainder of this paper, we model the macro-cell deployment according to a regular hexagonal network with an infinite number of cells while small-cells and their associated users are modeled according to dependent PPPs.\\

\subsubsection{Macro-cells deployment}

We consider a hexagonal cellular network denoted by $\Lambda$ with an infinite number of macro-cells having an intersite distance between them denoted by $\delta$. The hexagonal model means that for each node $\site$ $\in$ $\Lambda$, there exists a unique $(m,n)$ $\in$ $\mathbb{Z}^2$ such that $\site = \delta( m + ne^{i\frac{\pi}{3}})$. We denote by $\site_0$ the serving cell located at the origin of $\mathbb{R}^2$ ($\mathbb{R}^2$ is isomorphic to $\mathbb{C}$). Antenna in each site is assumed to have an omni-directional radiation pattern and covers a geographical area named Voronoi cell, having a cell radius denoted by $R$. Furthermore, the location of a mobile served by $\site_{0}$ is denoted by $z_0$ such that $z_0 = re^{i\theta}$ where (r, $\theta$) are the polar coordinates in the complex plane. We denote also by $z$ the geographical location of a mobile served by a cell $\site$ $\in$ $\Lambda^{*}$ in the plane, where $\Lambda^{*}$ is the lattice $\Lambda$ without the serving cell $\site_0$. Location $z$ is written in the complex plane by $z = \site + \rho e^{i\phi}$, where $\rho$ and $\phi$ represents respectively the distance and the angle between $z$ and $\site$. Moreover, it is assumed that the locations of mobile $z$ in the plane are uniformly distributed.\\

\subsubsection{Small-cells deployment}

As we have mentioned previously,  stochastic point processes are practical to model the random distribution of small-cells in a dense urban environment. Most common approaches model DL cellular networks considering that  BSs are distributed according to a spatial PPP and users distributed uniformly in BS Voronoi cells \cite{andrews2011tractable}. For the UL transmission, it has been proposed in \cite{novlan2013analytical} an interesting approach considering a spatial PPP distribution of users in the plane with a uniform distribution of each BS in the Voronoi cell of the associated mobile. However, with D-TDD, for each transmission direction there is two sources of interference: BSs and mobiles. In order  to exploit the mathematical framework based on stochastic geometry and satisfy in the same time the D-TDD assumptions, we model the set of active mobiles $z$ served by the small-cells, denoted hereafter by $\tilde{\site}$,  by a PPP $\Phi$ of intensity $\lambda$. This implies that $z$ are uniformly distributed in the studied area.

Given that each small-cell $\tilde{\site}$ has one active mobile transmitting whether in DL, with probability $\alpha_{d}$,  or in UL with probability $\alpha_{u}$, we assume that each mobile is associated with the nearest small-cell. Furthermore, the position of each small-cell can be expressed in the complex plane by $\tilde{\site}=z+\rho_z e^{i\phi_z}$, with $(\rho_z,\phi_z)$ are the polar coordinates of $\tilde{\site}$ relatively to $z$. This means that also the set of small-cells forms a PPP obtained by a displacement of $\Phi$ (Displacement theorem \cite{andrews2016primer}), i.e., the process of mobiles and small-cells are two dependent PPPs. \\

Additionally, to reduce strong macro-cell  interference impact on users served by small-cells, especially in small-cells range expansion, we assume that FeICIC is implemented. With this feature, macro-cells reduce their transmitted power level during some specific sub-frames called Almost Blank Sub-frames (ABS) so that small-cells use those sub-frames to configure dynamically the UL-DL frame portions. Moreover, we assume that all macro-cells are well synchronized and adopt the same frame configurations i.e., S-TDD . This means that all the ABSs are assigned at the same time portion to all the small-cells to activate D-TDD and also macro-cell BSs interference can be neglected.

	\subsection{Propagation model}     
	
	To model the wireless channel, we consider the standard power-law path loss model based on the distance between a mobile $z$ and a BS $\site$ such that the path loss $L(\site,z)$ is given by
	
	\begin{equation}
	L(\site,z)=a|\site-z|^{2b},
	\label{pathloss}
	\end{equation}
	with $2b$ is the path loss exponent and $a$ is a propagation factor that depends on the type of the environment (indoor, outdoor...).\\
	
	Actually, characterizing the propagation in wireless channels is often performed through measurements and statistics from field experiments in different environments and under different conditions. Based on that, several mathematical formulations are obtained. Nevertheless, those models don't describe the real behavior of propagation in wireless channels. The path loss exponent is an important parameter that can have a tremendous effect on system performance. It refers to the rate of decay of power with respect to distance between a transmitting and receiving nodes. Path loss exponent values depend on the environment of propagation (outdoor, indoor...), the visibility between the transmitter and the receiver (Line of Sight, Non Line of Sight) and also the links between nodes (BS to BS, mobile to mobile, BS to mobile... ). In effect, the propagation environment between a BS and a mobile is not the same as the one between two mobiles. This latter is more dynamic and undergoes multiple reflexions and diffractions \cite{turkka2008path}.  Small values of the path loss exponent refers to favorable conditions for electromagnetic waves propagation (e.g., free space path loss exponent is always taken 2) while big values of path loss exponent refers to harsh propagation conditions. In this paper, we choose to fix the same value of the path loss exponent for all  the propagation directions between transmitting and receiving nodes in order to alleviate notations (since we are dealing with equations having a lot of parameters). Meanwhile, the model can easily be adapted by taking a convenient path loss exponent for each propagation link. \\   
	
	In addition to the path loss, the received power by a mobile depends on the random channel effects, especially shadowing and fast fading. Shadowing refers to the attenuation of the received signal power caused by obstacles obstructing the propagation between the transmitter and receiver. The common approach to model shadowing effect in in cellular networks is to consider a sequence of independent log-Normal random variables multiplied by the expression of the path loss provided previously. In this work, shadowing effect is not considered to keep the tractability of our models when calculating the coverage probability. Meanwhile, we show in the following section how log-Normal shadowing can be included in the average $ISR$ derivation.\\

	Likewise, fast fading random model is not considered for macro-cell deployment analysis in order to simplify calculations. Actually, fading effect can be compensated through link level performing that maps the $SINR$ to the throughput ($Th$). Also, for an AWGN (Additive Gaussian Noise Channel), Shannon's formula provides the relation between $SINR$ and $Th$. Hence, the fast fading effect can be compensated by using a modified Shannon's formula to have $Th=K_1 log_2(1+K_2 SINR)$, with $K_1$ and $K_2$ are constants calibrated from practical systems \cite{mogensen2007lte}. However, for the heterogeneous system analysis, multi-path Rayleigh fading effect is considered. We denote by $H_i$ and $G_i$, with $i=\tilde{\site}$ or $i=z$, the fading coefficients between, a transmitting node $i$ and a typical receiving mobile, and between $i$ and a typical receiving BS. We assume also that the RVs $H_{i}$ and $G_{i}$ are independent and identically distributed (i.i.d) for each propagation link and follow an exponential distribution of mean 1.\\
	
	For the UL transmission, power control is applied to the Physical Uplink Shared Channel (PUSCH) in order to set the required mobile transmitted power. In this paper, it is modeled by the fractional power control model (FPC), i.e., the path loss is partially compensated by the power control \cite{castellanos2008performance}. The transmitted power by the mobile location $z$ to its serving cell $\site$ is then written
	
	\begin{equation}
	P(z,\site)= P^*(\site) \left(\left|z - \site \right|^{2b}\right)^{k},
	\label{power control}
	\end{equation}
	where $P^*$($\site$) is the cell specific target power and $k$ $\in$ [0,1] is the power control compensation factor. When $k=1$ the power control scheme totally indemnifies the path loss in order to reach the target power $P^*(\site)$. For the case $0<k<1$ the path loss is partially compensated and mobile users in cell edge create less interference because their transmitted power is reduced. \\
	
	Without loss of generality, we consider that $P^*(\site)$ is the same for all the cells. Power values $P$ and $P^*$ are supposed to include the path loss constants and antenna gains of BSs and user equipments. It is important to note that section 3 and section 4 are independent. So, to avoid confusion, we denote by $\tilde{P}$ and $\tilde{P^*}$ respectively the transmitting power and the cell specific target power of small-cell BSs.  
	
	\section{Dynamic TDD interference derivation in a macro-cell deployment}
	
	We define the Interference to Signal Ratio $ISR$ in DL as the received power from an interfering source (interfering mobile or BS) divided by the useful power received by $z_0$ from the serving cell. The average $ISR$ experienced in DL transmission by a mobile location $z_0$ connected to $\site_0$ is
	
	\begin{equation}
	D(z_0)= \alpha_{d}~D_{\downarrow}(z_0) + \alpha_{u}~D_{\uparrow}(z_0)
	\label{ISRtotalDL}
	\end{equation}
	with $D_{\downarrow}$ and $D_{\uparrow}$ are respectively DL to DL and UL to DL average $ISRs$ experienced during the DL cycle by $z_0$.\\
	
	Likewise, the average $ISR$ experienced by a cell $\site_0$ in UL transmission cycle is defined by
	\begin{equation}
	U(z_0)= \alpha_{u}~U_{\uparrow}(z_0) + \alpha_{d}~U_{\downarrow}(z_0)
	\label{ISRtotalUL}
	\end{equation}
	where $U_{\uparrow}$ and $U_{\downarrow}$ are respectively UL to UL and DL to UL interference to signal ratios experienced during the UL cycle of $\site_0$.
	
	\subsection{Downlink $ISR$ derivation $D(z_0)$}
	
	\subsubsection{Expression of DL to DL $ISR$ $D_{\downarrow}(z_0)$}
	
	In \cite{nasri2016Analytical}, it has been shown that the DL $ISR$ function of a location $z_0 = re^{i\theta}$ in a hexagonal cellular network with infinite number of cells admits a series expansion on $r$ and $\theta$ and is a very slowly varying function on $\theta$. Taking $x=\frac{r}{\delta}$ such that $x<1$ (for hexagonal networks, we always have $x<\frac{1}{\sqrt{3}}$), the expression of $D_{\downarrow}$ is recalled from \cite{nasri2016Analytical}
	\begin{equation}
	D_{\downarrow}(z_0) = \frac {6x^{2b}}{\Gamma(b)^{2}}\sum_{h=0}^{+\infty}\frac{\Gamma(b+h)^{2}}{\Gamma(h+1)^{2}}\omega(b+h)x^{2h}
	\label{ISRdlm}
	\end{equation}
	where $\Gamma(.)$ is the Euler Gamma function  and
	\begin{equation}
	\omega(z) = 3^{-z}\zeta(z)\left( \zeta(z,\frac{1}{3})-\zeta(z,\frac{2}{3})\right),
	\label{omega}
	\end{equation}
	with $\zeta(.)$ and $\zeta(.,.)$ are respectively the Riemann Zeta and Hurwitz Riemann Zeta functions \cite{abramowitz1964handbook}.\\
	
	\subsubsection{Expression of UL to DL $ISR$ $D_{\uparrow}(z_0)$}
	
	The UL to DL interference is generated from mobile users located at other cells, mainly from those located at the border of cells adjacent to the serving cell $\site_0$. Since there is only one mobile user transmitting at the same time in UL for each cell, the total UL to DL $ISR$ can be evaluated by averaging over locations $z\in\site$ and then summing over $\site\in$ $\Lambda^{*}$. So, if we assume that location $z$ is uniformly distributed in $\site$, $D_{\uparrow}(z_0)$ is mathematically written as
	
	\begin{equation}
	D_{\uparrow}(z_0)= \frac{1}{\pi R^{2}}\int \limits_{0}^{R} \int \limits_{0}^{2 \pi}\sum_{\site \in \Lambda^{*}}\frac {P^{*}~\rho^{2bk}~r^{2b}}{P\left|\site + \rho e^{i\phi}- r e^{i\theta} \right|^{2b}}\rho d\rho d\varphi
	\label{ISRultodl}
	\end{equation}
	
	To evaluate equation (\ref{ISRultodl}), we can proceed analogously to the proof of $ISR$ formulas in hexagonal omni-directional networks provided in \cite{nasri2016Analytical}. We start by taking $z^{'}= re^{i\theta} - \rho e^{i\phi}$. It is obvious that $\left|z^{'}\right| < \left|\site \right|$. It follows from \cite{nasri2016Analytical} that the sum over $\site$ inside the double integral admits a series expansion on $\left|z^{'}\right|/\delta$ as in (\ref{ISRdlm}). Using formula (\ref{ISRdlm}) and writing $\left|z^{'}\right|$ in terms of $r$, $\theta$, $\rho$ and $\phi$, (\ref{ISRultodl}) becomes
	{
		
		\begin{align}
		D_{\uparrow}(z_0) &=\frac{6 P^{*}x^{2b}}{P \pi R^{2}~\Gamma(b)^{2}}\int \limits_{0}^{R} \int \limits_{0}^{2 \pi}\sum_{h=0}^{+\infty} \frac{\Gamma(b+h)^{2} \omega(b+h)}{\Gamma(1+h)^{2} \delta^{2h}} \times\nonumber\\
		&(r^{2} + \rho^{2} )^{h} (1- \frac{2r \rho}{r^{2} + \rho^{2}} \cos ( \phi ))^{h} \rho^{2bk+1} d\rho d\phi 
		\label{ISRultodl3}
		\end{align}
	}
	The sum and integrals of (\ref{ISRultodl3}) can be switched and the inside integral can be evaluated by expanding $(1- \frac{2r \rho}{r^{2} + \rho^{2}} \cos ( \phi ))^{h}$ as a binomial sum. After few derivations of known special integrals and simplifications, the UL to DL $ISR$ $D_{\uparrow}(z_0)$ can be evaluated by the following convergent series on $x=r/\delta$
	{	
		\begin{align}
		D_{\uparrow}(z_0) &= \frac{6 P^{*} x^{2b} R^{2bk}}{P~\Gamma(b)^{2}}\sum_{h=0}^{+\infty}\sum_{n=0}^{\lfloor \frac{h}{2} \rfloor}\sum_{i=0}^{h-2n} \frac{\Gamma(b+h)^{2} \omega(b+h)}{\Gamma(n+1)^{2}\Gamma(h+1)}\times\nonumber\\
		&\frac{ (\frac{R}{\delta})^{2n+2i}~x^{2h-2n-2i}}{\Gamma(i+1)\Gamma(h-2n-i+1) (n+i+bk+1)}
		\label{ultodlisrfinal1}
		\end{align}
	}
	Since $x < 1/\sqrt{3}$ for hexagonal model, it is obvious that the first elements of this series are sufficient to numerically evaluate $D_{\uparrow}$. Furthermore, after few simplifications, (\ref{ultodlisrfinal1}) can be written as an entire series on $x^{2}$ as follows
	
	\begin{equation}
	D_{\uparrow}(z_0)=\frac{6 P^{*} x^{2b} R^{2bk}}{P} \sum_{h=0}^{+\infty} \beta_h x^{2h}
	\label{uplinkISRxx}
	\end{equation}
	with
	
	\begin{align}
	\beta_h&=\sum_{n=0}^{h}\sum_{i=0}^{+\infty} \frac{\Gamma^2(b+h+n+i) \omega(b+h+n+i)}{\Gamma(b)^{2} \Gamma^2(n+1) \Gamma(i+1)} \times \nonumber \\
	& \frac{(\frac{R}{\delta})^{2n+2i}}{\Gamma(h-n+1) \Gamma(h+n+i+1) (n+i+bk+1)} 
	\label{betah}
	\end{align}
	
	\subsubsection{How to include shadowing in calculations}
	
	As we have mentioned previously, shadowing refers to the attenuation of the received signal power caused by obstacles obstructing the propagation between the transmitter and receiver. In general, shadowing between a transmitting node $t$ and a receiving node $r$ ($t$ and $r$ can be BSs or mobiles) is modeled by a log-Normal random variable $X_{t}(r)=10^{\frac{Y_{t}(r)}{10}}$ with $Y_{t}(r)$ is a Normal random variable with mean $\mathbb{E}(Y_{t}(r))=0$ and variance $\sigma^2$. \\
	
	To model the shadowing effect, we consider two independent and identically distributed sequence of log-Normal random variables $X_{\site}(z_0)$. The expression of the average DL to DL $ISR$ becomes
	
	\begin{equation}
	D_{\downarrow}(z_0) = \frac {6x^{2b}}{\Gamma(b)^{2}} \mathbb{E}[10^{\frac{\tilde{Y}_{\site}(z_0)}{10}}] \sum_{h=0}^{+\infty}\frac{\Gamma(b+h)^{2}}{\Gamma(h+1)^{2}}\omega(b+h)x^{2h}
	\label{ISRdlmshad}
	\end{equation}
	where $10^{\frac{\tilde{Y}_{\site}(z_0)}{10}}=\tilde{X}_{\site}(z_0) $ is a log-normal random variable representing the ratio of the shadowing effect from interfering cells and the shadowing effect from the serving cell (the ratio of two log-Normal random variables  is a log-Normal random variable), with $\tilde{Y}_{\site}$ is a Normal RV with mean 0 and variance $\tilde{\sigma}^2$. The same reasoning can be followed to include shadowing  when calculating the average UL to DL $ISR$ and  for the UL analysis provided in the next section.

	\subsection{Uplink $ISR$ derivation $U(z_0)$}
	
	In this part, we derive the analytical expression of the UL interference to signal ratio. The UL signal received from location $z_0$ at cell $\site_0$ experiences interference coming from cells transmitting in DL and also from mobiles in adjacent cells which are in UL transmission cycle. The following results may be proved in much the same way as $D_{\downarrow}$ and $ D_{\uparrow}$ in the previous section.\\
	
	\subsubsection{UL to UL $ISR$ $U_{\uparrow}(z_0)$}
	
	The UL interference is generated by mobiles in neighboring cells which are randomly distributed in the network as opposed to the DL direction where cells' positions are fixed. Thus recalling the fact that mobile location $z$ is uniformly distributed in cell $\site$ and taking into account the definition of the transmitted power with fractional power control model given by equation (\ref{power control}), $U_{\uparrow}(z_0)$ can be expressed as
	
	{\begin{align}
		U_{\uparrow}(z_0) =& \frac{1}{\pi R^2}\int \limits_{0}^{R} \int \limits_{0}^{2 \pi} \sum_{\site \in \Lambda^{*}} \frac{\rho^{2bk}~\left|\site+\rho~e^{i\phi}\right|^{-2b}}{r^{2b(k-1)}} \rho d\rho d\phi \nonumber\\
		=& A_{1}(b)~x^{2b(1-k)}
		\label{fds}
		\end{align}
	}
	where
	\begin{equation}
	A_{1}(b)=\frac{6(R/\delta)^{2bk}}{\Gamma(b)^{2}} \sum_{h=0}^{+\infty}\frac{\Gamma(b+h)^{2}~\omega(b+h)}{\Gamma(h+1)^{2}~(bk+h+1)}(R/\delta)^{2h}\nonumber
	\label{ddulinterference}
	\end{equation}
	
	It is interesting to note that when a mobile is located at the same position as the serving BS, the UL to DL interference expression becomes similar to the expression of the UL to UL interference. From equation(\ref{betah}) we have
	
	\begin{equation}
	\beta_0=\sum_{i=0}^{+\infty}\frac{\Gamma(b+i)^{2}~\omega(b+i)}{\Gamma(i+1)^{2}~(bk+i+1)}(R/\delta)^{2i}
	\label{beta0}
	\end{equation}
	which is similar to the expression of $\frac{1}{6} (\frac{\delta}{R})^{2bk} A_{1}(b)$.
	
	\subsubsection{DL to UL $ISR$ $U_{\downarrow}(z_0)$}
	
	The signal coming from neighboring cells is often very strong with respect to mobile transmit power, especially if neighboring cells' antennas are in LOS condition or inter-site distance is lower (path loss is low). Contrary to the UL to UL interference, here the interfering signals come from cells, which have fixed positions. Hence, under the same system model assumptions, $U_{\downarrow}$ is given by
	
	\begin{equation}
	U_{\downarrow}(z_0) = \sum_{\site \in \Lambda^{*}} \frac{P \left|\site\right|^{-2b}}{P^{*}~r^{2b(k-1)}} = A_{2}(b)x^{2b(1-k)}
	\label{dltoulinterference}
	\end{equation}
	where $A_{2}(b)=\frac{P~\omega(b)}{P^{*}~\delta^{2bk}}$.\\
	
	\begin{figure}[tb]
		\centering
		\includegraphics[height=7cm,width=9.5cm]{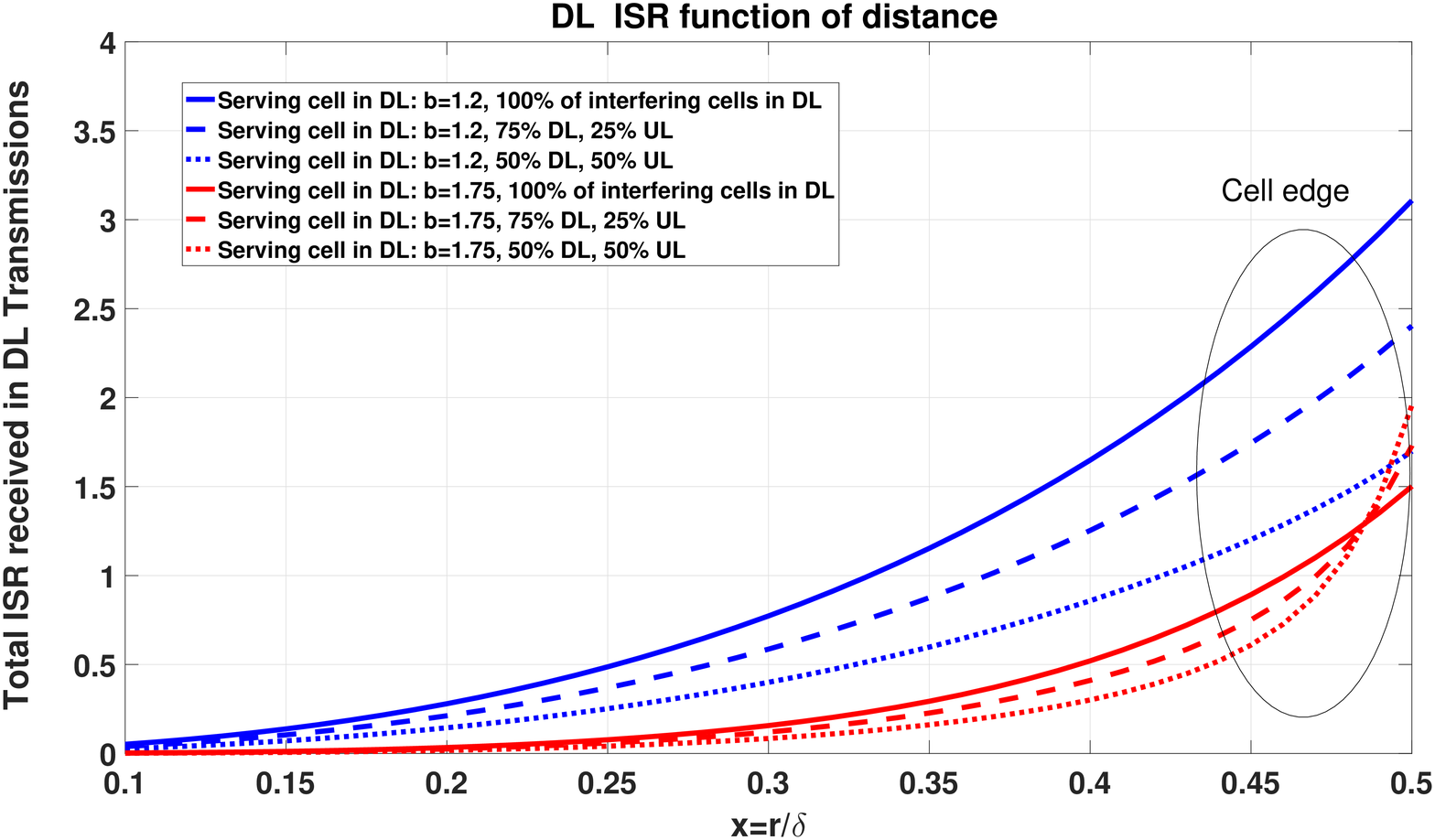}
		\caption{DL ISR: Static TDD vs Dynamic TDD.}
		\label{sim1}
	\end{figure}
	
	\begin{figure}[tb]
		\centering
		\includegraphics[height=7cm,width=9.5cm]{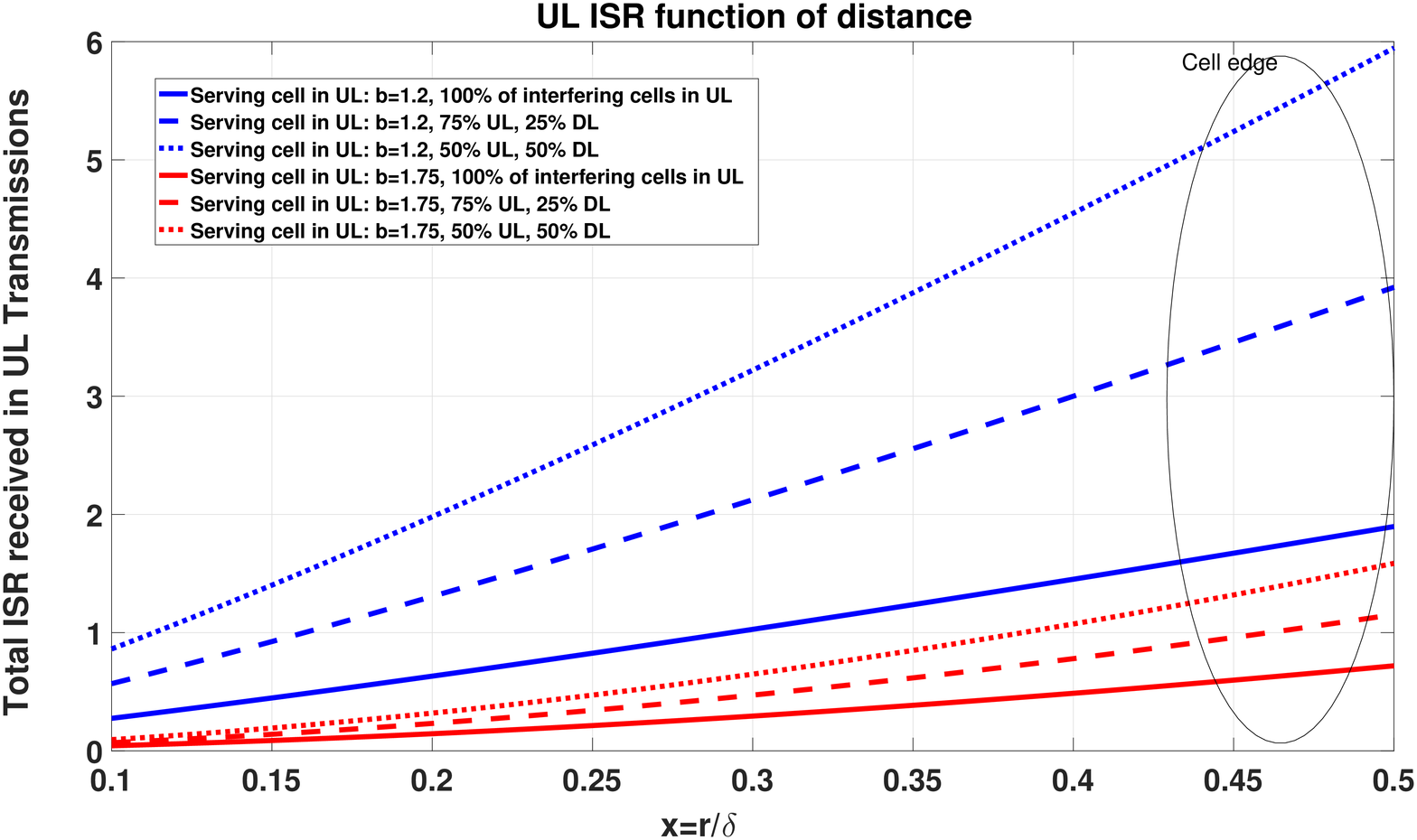}
		\caption{UL ISR: Static TDD vs Dynamic TDD.}
		\label{sim2}
	\end{figure}
	
	Fig. \ref{sim1} shows the developed $ISR$ in DL transmission direction for different values of path loss exponent (2b=2.4, 2b=3.5). The first obvious observation is that the DL interference level decreases in the studied cell when other cells use more frequently the UL transmission cycle. This means that the impact of DL interference coming from other cells is relatively higher than the impact of interference coming from mobiles. Consequently, one can conclude that DL interference level in DL cycle for D-TDD should be lower than Static TDD.\\
	
	The system behavior during the UL cycle is completely different. As shown in Fig. \ref{sim2}, interference level significantly increases when 25\% or 50\% of cells switched to the opposite direction, i.e., DL transmission. The UL performance degradation is mainly related to the higher DL transmit power of other cells, especially when they are in LOS conditions. This makes D-TDD system very limited by DL to UL interference. These conclusions are in agreement with the results of  \cite{khoryaev2012performance}, which showed that there is an improvement of 10dB in the DL $SINR$ of the serving cell when 50\% of other cells switch from DL to UL transmission cycle; whereas the UL $SINR$ of the same serving cell degrades by 20dB. This UL performance loss is expected to be more significant in macro-cell deployment. Therefore, DL to UL interference can seriously deteriorate system performance if no action is taken to mitigate it.
	
	\subsection{Coverage probability}
	
	The coverage probability (CCDF of $SINR$) is the probability that a mobile user is able to achieve a threshold $SINR$, denoted by $\gamma$, in UL and DL transmissions.
	
	\begin{equation}
	\Theta(\gamma) = P( SINR > \gamma)
	\label{coverage}
	\end{equation}
	
	For any scenario of user location distributions, the coverage probability is given by 
	
	\begin{equation}
	\Theta(\gamma) = \int_{\site_{0}} \mathds{1}( SINR > \gamma)dt(z)
	\label{distributioncov}
	\end{equation}
	such that $\int_{\site_{0}}dt(z) = 1$ (e.g., $dt(z)=\frac{rdrd\theta}{\pi R^{2}}$ for uniform user locations distribution).\\
	
	Based on the expressions of the DL and UL $ISR$ derived previously, we define the	DL and UL $SINR$, denoted respectively by $\Pi_{DL}$ and $\Pi_{UL}$, as follows
	
	\begin{align}
	&\Pi_{DL}(x) = \frac{1}{\eta~D(x) + y_{0} x^{2b}}=\frac{1}{d(x)} \\
	&\Pi_{UL}(x) = \frac{1}{\eta~U(x) + y^{'}_{0} x^{2b(1-k)}}=\frac{1}{u(x)}
	\label{SINRDL}
	\end{align}
	where $y_{0}=\frac{P_{N}\delta^{2b}}{P}$, $y^{'}_{0}=\frac{P_{N}\delta^{2b(1-k)}}{P^{*}}$, $P_{N}$ is the thermal noise power and $\eta$ is the average load over the interfering cells.\\

	Under a uniform user locations distribution, the expression of the coverage probability is given by
	
	\begin{equation}
	\Theta(\gamma) = \frac{2}{R^{2}}\int_{0}^{R}\mathds{1}( r < \delta g^{-1}(\frac{1}{\gamma}))~rdr=\frac{\varTheta^{2}(\gamma)}{R^2}
	\label{coverageproba}
	\end{equation}
	where 
	\begin{equation}
	\varTheta(\gamma)=\min \left( \delta \times g^{-1}(\frac{1}{\gamma}),R\right)\nonumber
	\end{equation}
	with $g=d$ for the DL coverage probability and $g=u$ for the UL coverage probability.\\
	
	The explicit formulas of the coverage probability require the inverse functions of $d$ and $u$. This can be calculated by using series reversion methods; see for instance \cite{nasri2016Analytical}. The inverse of $u$ is easy to derive, and it is given by
	
	\begin{equation}
	u^{-1}(y)= (\frac{y}{\eta \alpha_{u} A_{1}(b)  + \eta \alpha_{d} A_{2}(b) + y^{'}_{0}})^{\frac{1}{2b(1-k)}}
	\end{equation} 
	
	%The inverse function of $d$ requires extensive calculations and it is not provided here. Readers can refer to appendix D of \cite{nasri2016Analytical} to see a series reversion method that gives an approximation of $d^{-1}$.
	
	Now, to derive the inverse function of $d$, we shall follow the same approach as in \cite{nasri2016Analytical}. To do so, let $y=d(x)$, using the series expansion of $D_{\downarrow}(m)$ in (\ref{ISRdlm}) and the simplified expression of $D_{\uparrow}(m)$ given in (\ref{uplinkISRxx}), it is clear that $d$ admits an analytic expansion on $x=r/\delta$ and can be expressed as follows 
	\begin{equation}
	y=x^{2b} f(b) \left(1+ \sum_{h=1}^{+\infty}c_h x^{2h}\right)
	\label{gexpansion}
	\end{equation}
	where
	\begin{equation}
	f(b)=6\eta \alpha_d \omega(b)+ \frac{6\eta P^* R^{2bk} \beta_0}{P \Gamma^2(b)} + y_0
	\end{equation}
	and
	
	\begin{equation}
	c_h=\frac{6\eta \alpha_{d}\Gamma(b+h)^2\omega (b+h) \beta_h}{f(b) \Gamma(b)^2\Gamma(h+1)^2}
	\label{alphan}
	\end{equation}
	
	Equation (\ref{gexpansion}) can be transformed to
	\begin{align}
	\left(\frac{y}{f(b)}\right)^{\frac1b}&=x^{2}\left(1+\sum_{h=1}^{+\infty}c_h x^{2h}\right)^{\frac1b}\nonumber\\
	&=x^{2} + \frac{c_1}{b}x^{4}+O(x^{6})
	\label{gexpansion1}
	\end{align}
	
	Eliminating the error terms in equation (\ref{gexpansion1}) gives a second-order equation in $x^2$ that admits two solutions:
	
	\begin{align}
	x_{\pm}^{2}=\frac{V(y,b)^2}{\frac12\pm\sqrt{\frac14+ \frac{c_1}{b} V(y,b)^2}}
	\end{align}
	
	where 
	\begin{equation}
	V(y,b)=(\frac{y}{f(b)})^{\frac{1}{2b}}
	\end{equation}
	
	Since $x=\frac{r}{\delta}$ is a positive real number, only the positive solution is valid. Hence we obtain the following approximation of $x=d^{-1}(y)$
	
	\begin{equation}
	d^{-1}(y)\approx\frac{V(y,b)^{2}}{\sqrt{\frac12+\sqrt{\frac14+\frac{c_1}{b}V(y,b)^2}}}.
	\label{approxInvg}
	\end{equation}

	\section{Small-cell network performance analysis}
	
\begin{figure}[tb]
		\centering
		\includegraphics[height=7cm,width=9.5cm]{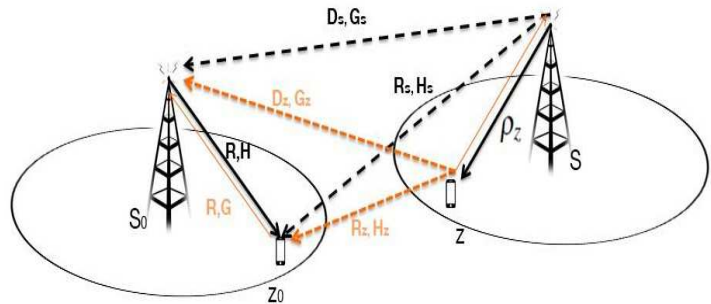}
		\caption{Parameters recapitulation.}
		\label{recapitul}
	\end{figure}

	Let $\tilde{\site}_0=z_0+Re^{i\theta}$ be the complex location of a typical small-cell that serves a typical mobile $z_0$. Thanks to the stationarity of the PPP $\Phi$, we can evaluate interference in the mobile location $z_0=0$ having a random distance $R$ to its closest serving BS. To simplify calculations, we denote hereafter by $R_{\site}$, the distance between an interfering small-cell $\tilde{\site}$ and the typical mobile location $z_0$, by $R_z$ the distance between an interfering mobile $z$ and $z_0$, by $D_{\site}$ the distance between an interfering small-cell $\tilde{\site}$ and the typical cell $\tilde{\site}_0$ and by $D_z$ the distance between an interfering mobile $z$ and $\tilde{\site}_0$; see Fig. \ref{recapitul}.\\
	
	Based on the heterogeneous system model provided previously, The  DL interference $\mathcal{I}_{DL}$, perceived by a mobile operating in DL, is expressed by
	{
		\begin{align}
		&\mathcal{I}_{DL} = \mathcal{I}_{DL->DL} + \mathcal{I}_{UL->DL} \nonumber \\
		&=\sum_{z \in \Phi\setminus\{z_0\}} \big( R^{-2b}_{\site} H_{\site} \tilde{P} \chi_{d}(z) +  R^{-2b}_z H_{z}\rho^{2bk}_z \tilde{P}^* (1-\chi_d(z))  \big). 
		\label{DLI}
		\end{align}
	} 
	
	Similarly, the  UL interference $\mathcal{I}_{UL} $ experienced by a mobile $z_0$, operating in UL transmission cycle and received at the serving small-cell BSs position, is given by
	
	{
		\begin{align}
		&\mathcal{I}_{UL} = \mathcal{I}_{UL->UL} + \mathcal{I}_{DL->UL} \nonumber \\
		&=\sum_{z \in \Phi\setminus\{z_0\}} \big( D^{-2b}_z  G_{z} \rho^{2bk}_z \tilde{P}^* \chi_u(z) + D^{-2b}_{\site}  G_{\site} \tilde{P} (1-\chi_u(z))  \big).
		\label{ULI}
		\end{align}
	} 
	
	Therefore, the DL and UL $SINR$ can be defined respectively by
	
	\begin{align}
	&\Pi_{DL}= \frac{\tilde{P} H R^{-2b}}{\mathcal{I}_{DL}+ P_N}  \\
	&\Pi_{UL}=\frac{\tilde{P}^* G R^{-2b(1-k)} }{\mathcal{I}_{UL}+ P_N}.
	\label{SNR}
	\end{align}
	with $H$ and $G$ are the fading coefficients between the typical mobile and its serving small-cell.\\
	
	Once again, the DL (UL) coverage probability is defined as the CCDF of the DL (UL) $SINR$. It gives the percentage of locations in which $\Pi_{DL}$ ($\Pi_{UL}$) is greater than a threshold value $\gamma$. It can be expressed for the DL and UL transmission directions as
	
	\begin{align}
	&\Theta_{DL}= \mathbb{E}_{\{R\}} \big[ \mathbb{P}(\Pi_{DL}> \gamma)|R=r \big] \\
	&\Theta_{UL}=\mathbb{E}_{\{R\}} \big[ \mathbb{P}(\Pi_{UL}>\gamma)|R=r \big] .
	\end{align}

	\subsection{DL coverage probability derivation} 
	
	Starting from the definition of the DL coverage probability and the DL $SINR$, we have
	
	\begin{equation}
	\Theta_{DL}(\gamma) = \int_{0}^{+\infty} \mathbb{P}\big[ H > \frac{\gamma (\mathcal{I}_{DL} + P_N)}{\tilde{P} R^{-2b}}| R=r \big] f_R(r) dr
	\label{dlcoverage1}
	\end{equation}  
	with $f_R(r)$ is the distribution of $R$ which is the distance between $z_0$ and the closest small-cell. Using the null probability of a PPP, it has been shown that this distance is Rayleigh distributed \cite{andrews2011tractable} and its probability density function is given by 
	
	\begin{equation}
	f_R(r)=2\pi \lambda r e^{-\lambda \pi r^2}
	\label{closestR}
	\end{equation}
	
	Using the fact that $H$ follows an exponential distribution of mean 1 and the definition of the Laplace transform, it follows that
	
	\begin{equation}
	\Theta_{DL}(\gamma) = 2\pi \lambda \int_{0}^{+\infty} e^{-\lambda \pi r^2}  e^{-\gamma \tilde{P}^{-1} P_N r^{2b }} \mathcal{L}_{\mathcal{I}_{DL}}(\frac{\gamma r^{2b }}{\tilde{P}} ) rdr
	\label{dlcoverage2}
	\end{equation}
	with $\mathcal{L}_{\mathcal{I}_{DL}}(v)$ is the Laplace transform of $\mathcal{I}_{DL}$ conditionally on $R$. It is defined by
	
	\begin{align}
	&\mathcal{L}_{\mathcal{I}_{DL}}(v)=\mathbb{E}\big[ e^{-v {\mathcal{I}_{DL}} } \big] \nonumber \\
	&=\mathbb{E}_{\{R_{\site},R_z,\rho_z,H_{\site},H_z, \chi_{d}(z)\}}\bigg[ \exp\big(-v \times \nonumber \\
	& \sum_{z \in \Phi\setminus\{z_0\}}  R^{-2b}_{\site} H_{\site} \tilde{P} \chi_{d}(z) +  R^{-2b}_z H_{z}\rho^{2bk}_z \tilde{P}^* (1-\chi_{d}(z)) \big) \bigg]
	\label{DLlaplace1}
	\end{align}
	
	From the complex geometry, the distance $R_{\site}$ between a small-cell $\tilde{\site}$ and the typical mobile $z_0=0$ can be written in terms of $R_z$ and $\rho_z$ as
	
	\begin{equation}
	R^2_{\site}= R^2_z + \rho^2_z + 2 R_z \rho_z \cos(\arg(z)-\phi_z).
	\end{equation}
	with $\arg(z)$ is the complex argument of $z$ relatively to the origin of the plane $z_0=0$.
	
	Thus, replacing $R_{\site}$ by its expression and using the fact that $H_{\site}$ and $H_z$ are i.i.d RVs and follow an exponential distribution of mean 1, (\ref{DLlaplace1}) can be simplified to
	
	\begin{align}
	&\mathcal{L}_{\mathcal{I}_{DL}}(v)=\mathbb{E}_{\{R_z,\rho_z,\chi_{d}(z)\}}\bigg[ \prod_{z \in \Phi\setminus\{z_0\}} \mathbb{E}_{\{H_{\site},H_z \}} \big[   \nonumber \\ 
	& \exp\big(-v (R^2_z + \rho^2_z + 2 R_z \rho_z \cos(\arg(z)-\phi_z))^{-b} H_{\site} \tilde{P} \chi_{d}(z)\big) \times \nonumber \\
	&\exp\big(- v R^{-2b}_z H_{z}\rho^{2bk}_z \tilde{P}^* (1-\chi_{d}(z)) \big)\big]  \bigg] \nonumber \\
	&=\mathbb{E}_{\{ R_z\}} \bigg[ \prod_{z \in \Phi\setminus\{z_0\}} \mathbb{E}_{\{\chi_{d}(z), \rho_z\}} \big[ \nonumber \\
	&\frac{1}{1+v (R^2_z + \rho^2_z + 2 R_z \rho_z \cos(\arg(z)-\phi_z))^{-b} \tilde{P} \chi_{d}(z)} \times \nonumber \\
	& \  \  \  \  \  \  \ \  \   \   \ \ \   \ \frac{1}{1+v R^{-2b}_z \rho^{2bk}_z \tilde{P}^* (1-\chi_{d}(z))} \big] \bigg]
	\label{DLlaplace2}
	\end{align}
	
	Now, using the Probability Generating Functional (PGFL) of PPP $\Phi$ with respect to the function inside the product, (\ref{DLlaplace2}) becomes
	
	\begin{align}
	&\mathcal{L}_{\mathcal{I}_{DL}}(v) = \exp\bigg( - \lambda \int_{r}^{+\infty} \int_{0}^{2\pi} \big(1- \mathbb{E}_{\{\chi_d(z), \rho_z\}} \big[ \nonumber \\
	& \frac{1}{1+v (x^2 + \rho^2_z + 2 x \rho_z \cos(\theta))^{-b} \tilde{P} \chi_{d}(z) } \times \nonumber \\
	&\frac{1}{1+v x^{-2b} \rho^{2bk}_z \tilde{P}^* (1-\chi_{d}(z))} \big]   \big) xdxd\theta  \bigg) \nonumber \\
	& =\exp\bigg( - \lambda \int_{r}^{+\infty} \int_{0}^{2\pi} \big(1- \mathbb{E}_{\{\rho_z\}} \big[ \nonumber \\
	& \frac{\alpha_{d}}{1+v (x^2 + \rho^2_z + 2 x \rho_z \cos(\theta))^{-b} \tilde{P} } + \nonumber \\
	&\frac{\alpha_{u}}{1+v x^{-2b} \rho^{2bk}_z \tilde{P}^* } \big]   \big) xdxd\theta  \bigg)
	\end{align}
	
	We have made the assumption that each user is associated with the nearest BS. Hence, following the same analysis provided in \cite{novlan2013analytical}, we can approximate the distribution of $\rho_z$ by the same distribution as $R$ i.e., Rayleigh. It follows that
	
	\begin{equation}
	f_{\rho_z}(\rho)=2\pi \lambda \rho e^{-\lambda \pi \rho^2},~~\rho \geq 0.
	\label{rhodistribution}
	\end{equation}
	
	Finally, by using (\ref{rhodistribution}), it follows that 
	
	\begin{align}
	&\mathcal{L}_{\mathcal{I}_{DL}}(v) = \exp\bigg( - \lambda \int_{r}^{+\infty} \int_{0}^{2\pi} \big(1- \nonumber \\ 
	&\int_{0}^{+\infty} 2\pi \lambda e^{-\lambda \pi \rho^2}\big[ \frac{\alpha_{d}}{1+v (x^2 + \rho^2 + 2 x \rho \cos(\theta))^{-b} \tilde{P} } + \nonumber \\
	&\frac{\alpha_u}{1+v x^{-2b} \rho^{2bk} \tilde{P}^*} \big] \rho d\rho   \big) xdxd\theta  \bigg).
	\end{align}

	\subsection{UL coverage probability derivation} 
	
	The derivation of the UL coverage probability is quite similar to the DL one. The typical cell $\tilde{\site}_0$ is operating in UL cycle and perceives interference coming from the other small-cells operating in DL and also from the mobiles transmitting in UL. Interference in this case is received at the location of $\tilde{\site}_0$. Since the small-cells are distributed also as a PPP, denoted hereafter by $\Phi_{\site}$ and constructed by the displacement of $\Phi$, we can perform the analysis at the location $\tilde{\site}_0=0$ (The stationarity of the PPP). Therefore, using the definition of the UL coverage probability given by (3.33), we get
	
	\begin{align}
	\Theta_{UL}(\gamma) &=\int_{0}^{+\infty} \mathbb{P}\big[ G > \frac{\gamma (\mathcal{I}_{UL} + P_N)}{\tilde{P}^* R^{-2b(1-k)}}| R=r \big] f_R(r) dr \nonumber \\
	&= 2\pi \lambda  \int_{0}^{+\infty} e^{-\lambda \pi r^2}  e^{-\gamma \tilde{P}^{*-1} P_N r^{2b(1-k) }} \times \nonumber \\      &\     \  \  \ \ \   \ \ \ \ \ \ \ \mathcal{L}_{\mathcal{I}_{UL}}(\frac{\gamma r^{2b(1-k) }}{\tilde{P}^*} ) rdr,
	\end{align} 
	with $\mathcal{L}_{\mathcal{I}_{UL}}(v)$ is the Laplace transform of $\mathcal{I}_{UL}$ calculated conditionally on $R$.
	
	Once again, the distance $D_{\site}$ represents the distance between an interfering small-cell $\tilde{\site}$ and the typical one taken at the origin ($\tilde{\site}_0=0$). Hence using the complex notations, we get
	
	\begin{equation}
	D^2_{z}= D^2_{\site} + \rho^2_z + 2 D_{\site} \rho_z \cos(\arg(\site)-\phi_z).
	\end{equation}
	
	Thus, by following the same steps as we did in the derivation of $\mathcal{L}_{\mathcal{I}_{DL}}(v)$, the Laplace transform of $\mathcal{I}_{UL}$ is given by
	
	%\begin{align}
	%&\mathcal{L}_{\mathcal{I}_{UL}}(v)= \alpha_u \exp\bigg( - \lambda \int_{r}^{+\infty} \int_{0}^{2\pi} \big(1- \nonumber \\ 
	%&\int_{0}^{+\infty} \big[ \frac{2\pi \lambda e^{-\lambda \pi \rho^2}}{1+v \rho^{2bk} (x^2 + \rho^2 - 2 x \rho \cos(\theta))^{-b} \tilde{P}^*} \big] \rho d\rho   \big) xdxd\theta  \bigg) + \nonumber \\
	%& \alpha_d \exp\bigg( - \lambda \int_{r}^{+\infty} \int_{0}^{2\pi} \big(1- \int_{0}^{+\infty} \big[ \frac{2\pi \lambda e^{-\lambda \pi \rho^2}}{1+v x^{-2b} \tilde{P}} \big] \rho d\rho   \big) xdxd\theta  \bigg)  .
	%\end{align} 
	
	\begin{align}
	&\mathcal{L}_{\mathcal{I}_{UL}}(v) = \exp\bigg( - \lambda \int_{r}^{+\infty} \int_{0}^{2\pi} \big(1- \nonumber \\ 
	&\int_{0}^{+\infty} 2\pi \lambda e^{-\lambda \pi \rho^2}\big[ \frac{\alpha_{u}}{1+v \rho^{2bk} (x^2 + \rho^2 - 2 x \rho \cos(\theta))^{-b} \tilde{P}^*} + \nonumber \\
	&\frac{\alpha_d}{1+v x^{-2b} \tilde{P}} \big] \rho d\rho   \big) xdxd\theta  \bigg).
	\end{align}
	
	\subsection{Average spectral efficiency}
	
	The instantaneous spectral efficiency (SE) is defined as the maximum information rate that can be transmitted in a given bandwidth. Using the upper-bound of the well known Shannon's formula, the instantaneous spectral efficiency is expressed by
	
	\begin{equation}
	SE_s=\log_2(1+\Pi_{s})
	\label{SE}
	\end{equation}
	with $s=DL$ when the serving small-cell is operating in DL and $s=UL$ when the serving small-cell is operating in UL.
	
	The average spectral efficiency (ASE) is obtained by averaging over (\ref{SE}). It follows that 
	
	\begin{align}
	ASE_s&=\mathbb{E}[\log_2(1+\Pi_{s}) ] \nonumber \\
	&=\int_{0}^{+\infty} \mathbb{P}(\log_2(1+\Pi_s) > t) dt \nonumber \\
	& \myeq \frac{1}{\ln(2)} \int_{0}^{+\infty}\frac{\Theta_{s}(\gamma)}{(\gamma+1)} d\gamma
	\end{align} 
	with $(a)$ comes from the change of variable $\gamma=e^t-1$. \\

	\section{Simulation results}
	
	We simulate in MATLAB the proposed Macro-cell and small-cell models, for both DL and UL transmission directions, using the parameters summarized in Table. \ref{tab:table1}.\\
	
	\begin{table}[h!]
		\begin{center}

			\begin{tabular}{l|c|r} % <-- Alignments: 1st column left, 2nd middle and 3rd right, with vertical lines in between
				
				%	\textbf{Parameters} & \textbf{Values} \\

				\hline
				Macro-cells power $P$ & 60dBm \\
				
				\hline
				
				Small cells power $\tilde{P}$ & 26dBm  \\ 
				
				\hline 
				Target power cell specific  $P^*$ & 20dBm \\ 
				
				\hline 
				Noise power  $P_N$ & -93dBm \\ 
				
				\hline
				Number of rings (Macro-cells)  & 4 (60 interfering BSs) \\
				
				\hline
				Inter-site distance $\delta$ & 1km   \\
				
				\hline 
				Antennas gain & 16dBi \\ 
				
				\hline
				User distributions & uniform   \\
				
				\hline
				small-cells density  & 10 cells.$km^{-2}$   \\
				
				\hline
				Propagation factor $a$ & Outdoor: 130dB, Indoor: 160dB    \\
				\hline
				System bandwidth & Macro:20Mhz, small:10Mhz  \\
				\hline
				
				Path loss exponent $2b$ & 2.5, 3.5  \\
				\hline
				
			\end{tabular}
			\caption{Simulation parameters.}
			\label{tab:table1}
		\end{center}
	\end{table}

\begin{figure}[tb]
		\centering
		\includegraphics[height=7cm,width=9.5cm]{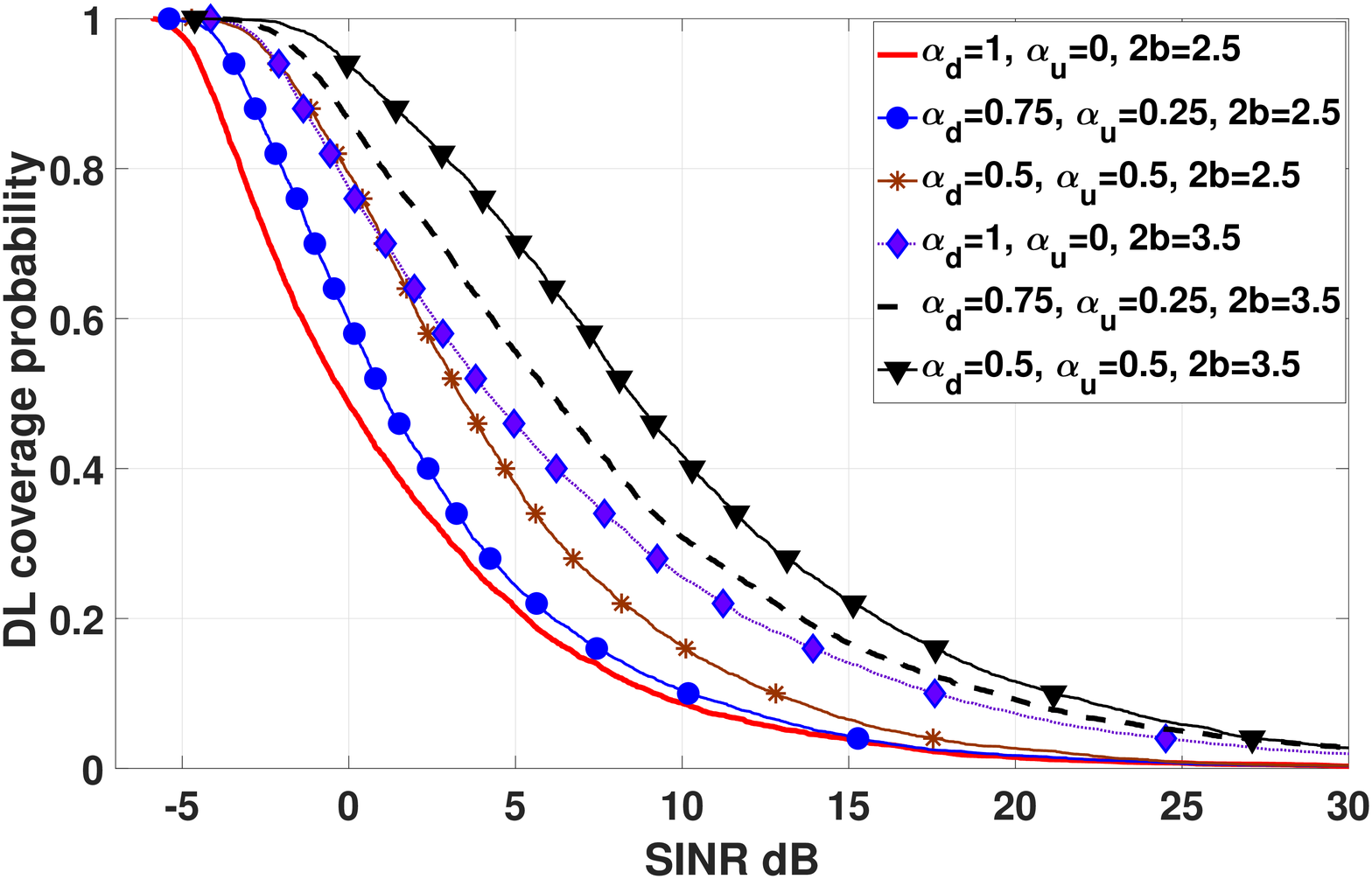}
		\caption{Macro-cells network DL coverage probability.}
		\label{dl1}
	\end{figure}

\begin{figure}[tb]
		\centering
		\includegraphics[height=7cm,width=9.5cm]{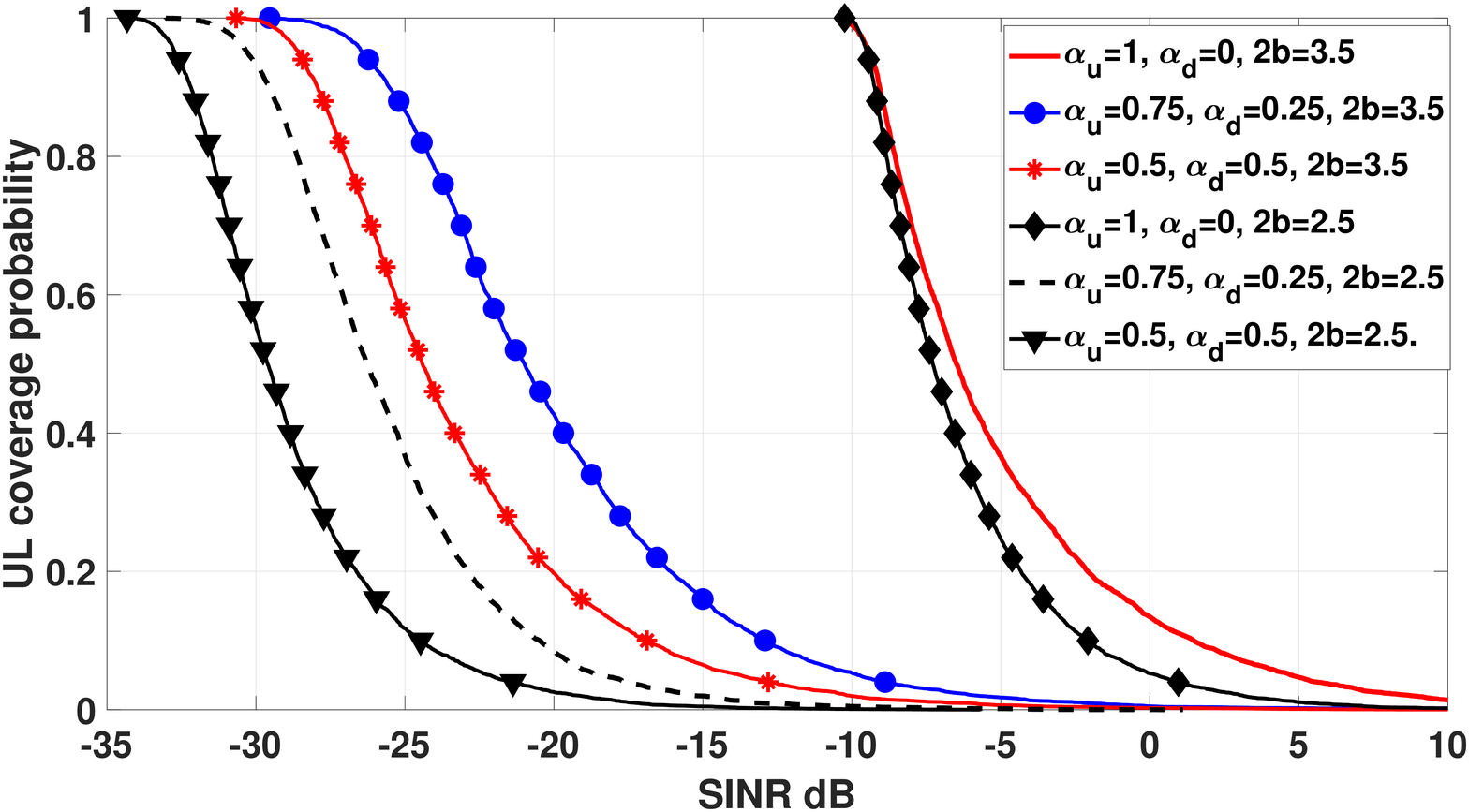}
		\caption{Macro-cells network UL coverage probability.}
		\label{ul1}
	\end{figure}

	We plot,  respectively in Fig. \ref{dl1} and Fig. \ref{ul1}, the DL and UL empirical coverage probability curves obtained by using Monte Carlo simulations for 20000 mobile locations $z_0$ and for two path loss exponent different values ($2b=2.5$ and $2b=3.5$). As we can see in Fig. \ref{dl1}, starting from a static TDD configuration where all the macro-cells are transmitting in DL, the coverage probability increases when D-TDD is activated with $\alpha_{d}=75\%$ and $\alpha_{d}=50\%$. This behavior is expected since the macro-cells BSs transmit with high power level and generate strong interference compared to interfering mobiles $z$ transmitting in UL. However, the system behavior is completely different during the UL cycle of the serving cell. As it is shown in Fig. \ref{ul1}, activating D-TDD with a mean number of neighboring UL macro-cells  $\alpha_{u}=75\%$ and $\alpha_u=50\%$ deteriorates completely the coverage probability. For instance, with a threshold SINR of $-20dB$, the coverage probability undergoes a huge degradation of $80\%$. Hence, one can conclude that D-TDD has a tremendous effect on performance during UL transmission especially for macro-cells deployment. Those results are in agreement with simulation results provided in \cite{khoryaev2012performance} and quite similar to the behavior of theoretical $ISR$ curves of Fig. \ref{sim1} and Fig. \ref{sim2}.\\

\begin{figure}[tb]
		\centering
		\includegraphics[height=7cm,width=9.5cm]{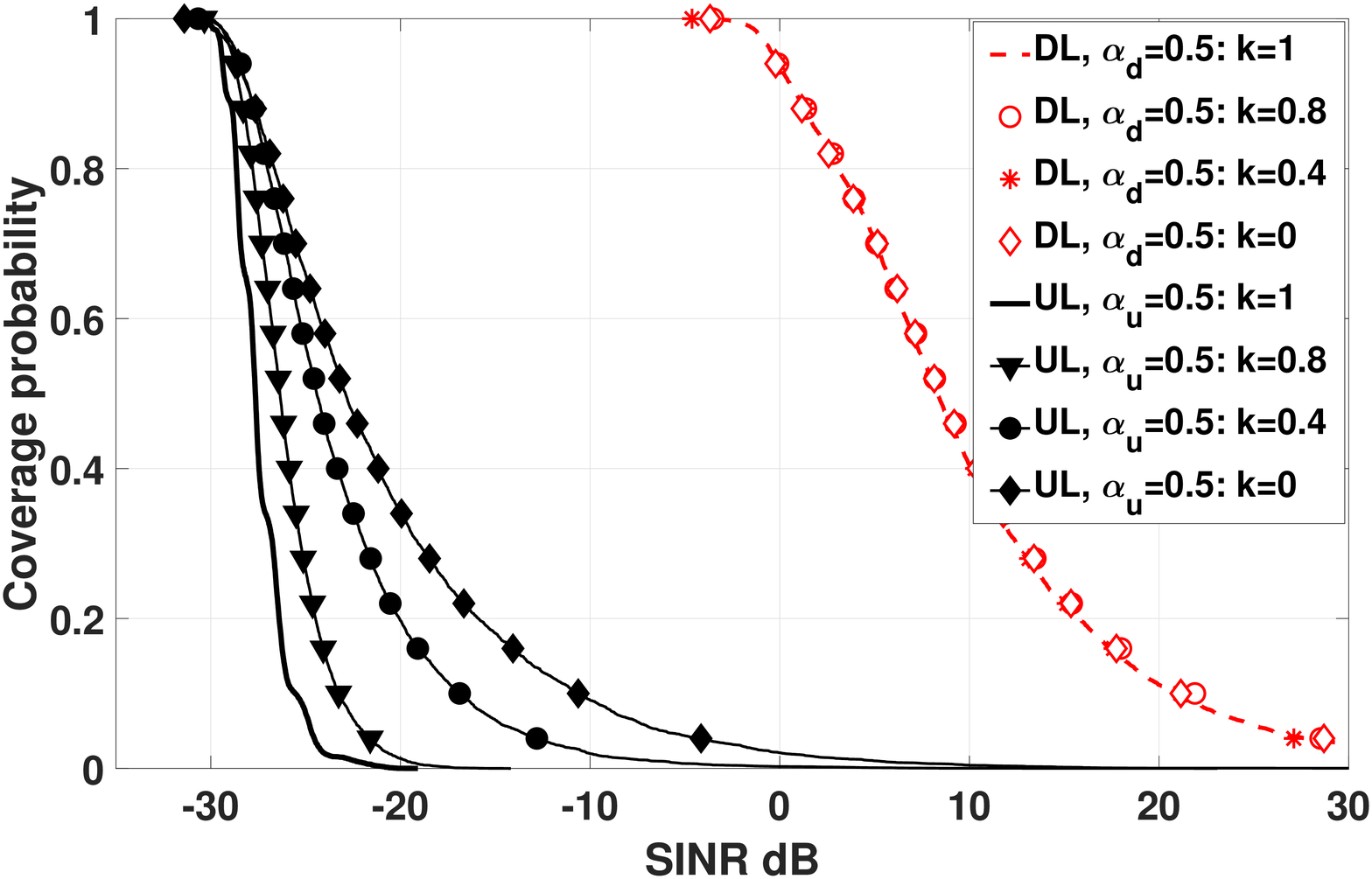}
		\caption{Macro-cells network DL and UL coverage probability with different power control factors.}
		\label{pc1}
	\end{figure}

	In Fig. \ref{pc1}, we plot the DL and UL coverage probability curves, with different fractional power control factor values ($k=0$, $k=0.4$, $k=0.8$ and $k=1$). We consider the scenario when D-TDD is activated and the number of DL and UL interfering cells is quite proportional (i.e., $\alpha_{d}=50\%$ and $\alpha_u=50\%$). When the serving cell is operating in DL, we notice that changing the power control factor has no impact on the coverage probability. This is mainly due to the fact that the principal interference impact comes from the DL BS signals where no power control mechanisms are considered. During the UL transmission cycle of the serving cell, one can notice that the coverage probability is decreasing when the fractional power control is increasing. Actually, FPC aims at providing the required $SINR$ to UL users while controlling at the same time their interference. When FPC factor $k=1$ the path loss is completely compensated and the target cell-specific power $P^*$ is reached. Thus, the interference coming from mobiles $z$ in UL is higher especially if a mobile is located in the edge of a neighboring cell. When FPC factor $0<k<1$, the scheme indemnifies partially the path loss. The higher is the path loss the lower is the received signal. This means that there is a compromise between the path loss and the $SINR$ requirements. Therefore, interference are likely to be controlled, which explain the enhancement of the coverage probability. This enhancement is more obvious when $k=0$. Which means that there is no compensation and the signal coming from the mobiles is weak.\\
	
	\begin{figure}[tb]
		\centering
		\includegraphics[height=7cm,width=9.5cm]{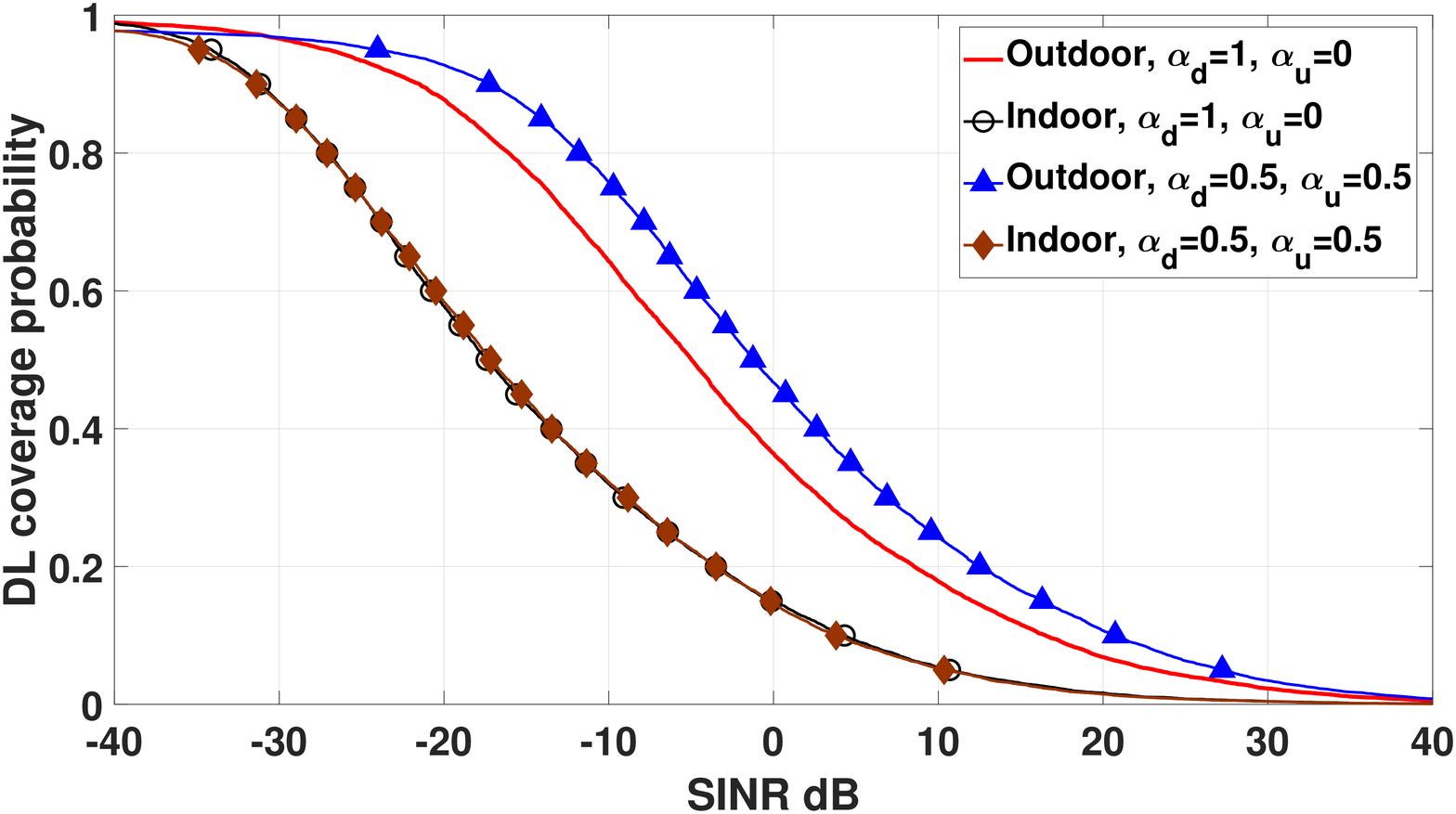}
		\caption{Small-cells DL coverage probability : $2b=3.5$, $k=0.4$ and $\lambda=10$ small-cell$/km^2$.}
		\label{ppp1}
	\end{figure}
	
	\begin{figure}[tb]
		\centering
		\includegraphics[height=7cm,width=9.5cm]{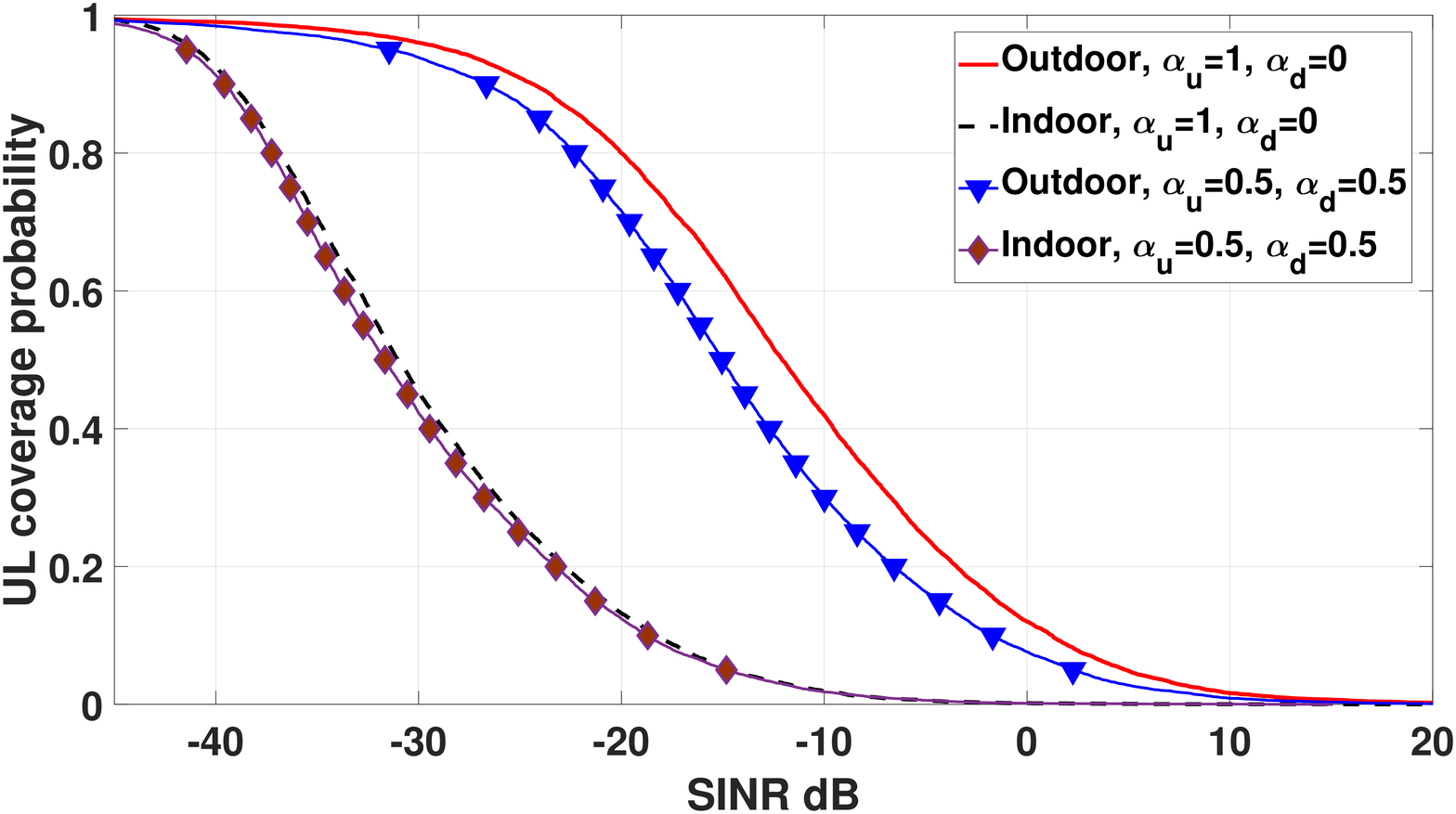}
		\caption{Small-cells UL coverage probability : $2b=3.5$, $k=0.4$ and $\lambda=10$ small-cell$/km^2$.}
		\label{ppp2}
	\end{figure}
	
	Fig. \ref{ppp1} shows the DL coverage probability obtained from the simulation of a heterogeneous network. First, small-cells operate with a static TDD configuration i.e., $\alpha_d=1$, then the whole network switch to a D-TDD configuration during the ABS sub-frames. We plot the coverage probability curves for an outdoor environment with a propagation parameter $a=130dB$ and a deep indoor environment with $a=160dB$. For the outdoor environment, the comparison between the S-TDD and D-TDD shows that the behavior is quite similar to macro-cells deployment results. For instance, when $50\%$ of small-cells switch from the DL to UL, there is an enhancement of the coverage probability by $15\%$ for a threshold $SINR$ of $-10dB$. However, for a deep indoor environment, one can notice that the coverage probability remains unchanged even when $50\%$ of the interfering small-cells switch to the opposite direction. In fact, signal propagation in a deep indoor environment suffers from high attenuation and delay factors because of the presence of obstacles and building penetration. During the DL transmission, the major interference comes from DL small-cells signal. In a deep indoor environment, not only the signal received from the serving cell is attenuated but also interference signal is subject to high attenuation. Similarly, Fig. \ref{ppp2} represents the comportment of the system during the UL transmission cycle. As expected, there is a degradation of the coverage probability when $50\%$ of small-cells switch from DL to UL for an outdoor environment. This degradation is not severe like the case of macro-cells because small-cells transmit with low power and not highly elevated. Also, the comportment in a deep indoor environment is quite similar to the DL scenario for the same reasons stated previously.\\

	\begin{figure}[tb]
		\centering
		\includegraphics[height=7cm,width=9.5cm]{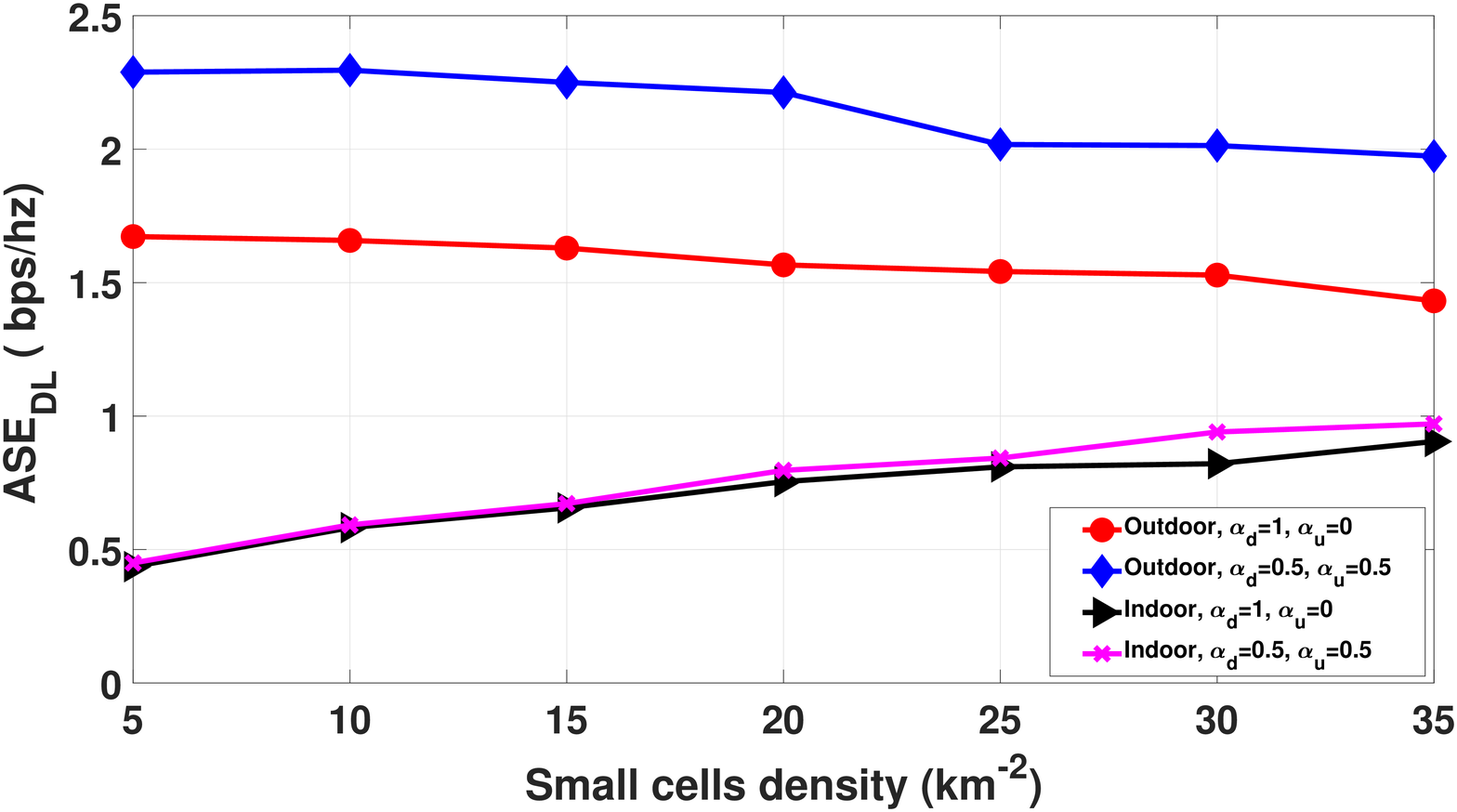}
		\caption{ DL average spectral efficiency : $2b=3.5$, $k=0.4$.}
		\label{ASE DL}
	\end{figure}
	
	\begin{figure}[tb]
		\centering
		\includegraphics[height=7cm,width=9.5cm]{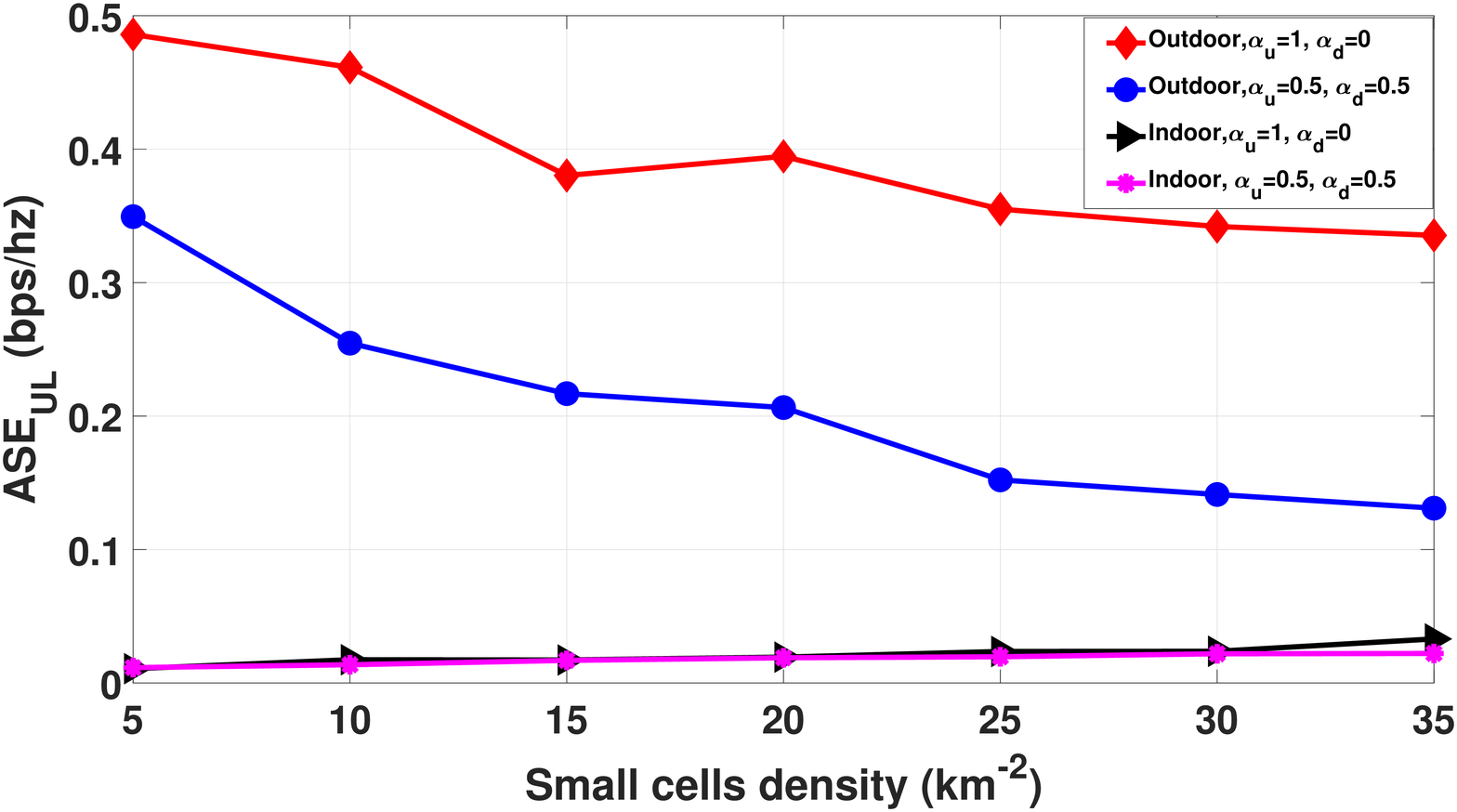}
		\caption{UL average spectral efficiency : $2b=3.5$, $k=0.4$..}
		\label{ASE UL}
	\end{figure}
	
	Additionally, we plot in Fig. \ref{ASE DL} the average spectral efficiency in DL obtained from the simulation of the HetNet as a function of small-cells' density. Once again, the comparison between the static TDD configuration and dynamic TDD, in an outdoor environment, shows that there is an enhancement of the ASE when D-TDD is activated with $\alpha_{d}=50\%$. Also, we observe that the ASE is slightly decreasing when small-cells density increases before it becomes almost constant. Actually, when small-cells density increases, the mean distance between the small-cells decreases because the density is inversely proportional to the mean distance between nodes. Hence, also the size of small-cells decreases and then both received signal power and interference increase simultaneously. Once we reach the interference limited scenario, the ASE becomes almost constant. For deep indoor environment, interference undergoes high attenuation as we have explained previously. When $\lambda$ increases, the size of small-cells decrease. Thus, the received signal from the serving cell overcome interference undergoing bad propagation conditions. Similarly in Fig. \ref{ASE UL}, we plot the ASE during the UL cycle of a typical serving small-cell using the same parameters as in DL. Once again, we observe that the ASE is decreasing when D-TDD is activated with $\alpha_{u}=50\%$ for the outdoor environment. Also, the ASE decreases as the small-cells density increases, especially when D-TDD is active. Moreover, for a deep indoor environment, the system experiences very bad performances in terms of ASE.\\

	\section{Conclusions}
	
	In this paper, we have investigated inter-cell interference in D-TDD based network. Explicit formulas of $ISR$ covering different scenarios of interference have been derived. We have provided the explicit expressions of the coverage probability in macro-cell and small-cell deployments. The comparison between static-TDD configuration and D-TDD shows that performance are better with D-TDD during the DL cycle of typical cell. However, the UL transmission is severely limited by interference coming from other BSs DL signals. Also, we have compared two types of environment, outdoor and deep indoor. As expected, the system experiences bad performance in deep indoor environment for both static and dynamic TDD. Moreover, we have analyzed the impact of fractional power control mechanisms on the UL transmission. Results have shown that small FPC factors enhance the coverage probability for both D-TDD and S-TDD. Further extension of this work could include the analysis of interference mitigation schemes such as 3D beamforming for macro-cell deployment in order to minimize the impact of the strong DL to UL interference which will make D-TDD feasible for macro-cells. Also, extension could include a dynamic system level analysis by including traffic model.     
	
	\bibliographystyle{IEEEtran}
	\bibliography{IEEEabrv,TelecomReferences}

\end{document}